\documentclass[9pt, conference]{IEEEtran}
\IEEEoverridecommandlockouts

\usepackage{cite}
\usepackage[table]{xcolor}  
\usepackage{multirow}
\usepackage{amsmath,amssymb,amsfonts}
\usepackage{algorithm}
\usepackage{algpseudocode}
\usepackage{algorithmicx}
\usepackage{graphicx}
\usepackage{textcomp}
\usepackage{booktabs} 
\usepackage{tikz}
\usepackage{tabularray}
\usepackage{booktabs}
\usepackage{float}
\usepackage{diagbox}
\usepackage{subcaption}
\usepackage{marvosym} 
\usepackage[hidelinks]{hyperref}
\usepackage[bottom=2.05cm, top=1.25cm,left=1.72cm,right=1.72cm]{geometry}

\newcommand{\circled}[1]{%
  \tikz[baseline=(char.base)]{
    \node[shape=circle,fill=black,draw,inner sep=1pt] (char) {\textcolor{white}{#1}};%
  }%
}
\def\BibTeX{{\rm B\kern-.05em{\sc i\kern-.025em b}\kern-.08em
    T\kern-.1667em\lower.7ex\hbox{E}\kern-.125emX}}
\begin{document}

\title{Conference Paper Title*\\
{\footnotesize \textsuperscript{*}Note: Sub-titles are not captured in Xplore and
should not be used}
\thanks{Identify applicable funding agency here. If none, delete this.}
}

\title{
BBAL: A \underline{B}idirectional \underline{B}lock Floating Point-Based Quantisation \underline{A}ccelerator for \underline{L}arge Language Models 

\author{\IEEEauthorblockN{\small Xiaomeng Han$^{1 \ast}$, Yuan Cheng$^{2,3 \ast \dag}$, Jing Wang$^{1}$, Junyang Lu$^{1}$, Hui Wang$^{1}$, X.x. Zhang$^{4}$, Ning Xu$^{1}$, Dawei Yang$^{2 \textrm{\Letter}}$, Zhe Jiang$^{1 \textrm{\Letter}}$}
\IEEEauthorblockA{
\textsuperscript{\rm 1}National Center of Technology Innovation for EDA, School of Integrated Circuits, Southeast University.
\textsuperscript{\rm 2}Houmo AI.\\
\textsuperscript{\rm 3}Nanjing University.
\textsuperscript{\rm 4}Jilin Normal University.\\
mingzhihan7@gmail.com,
yuancheng@smail.nju.edu.cn,
nick585108@163.com,
2972462961@qq.com,\\
whmio0115@seu.edu.cn,
3255594256@qq.com,
xning@seu.edu.cn,
dawei.yang@houmo.ai, 
zhejiang.uk@gmail.com
}
}
}

\maketitle
\def\thefootnote{$\textrm{\Letter}$}\footnotetext{Corresponding authors.}
\def\thefootnote{$\ast$}\footnotetext{Equal contribution.}
\def\thefootnote{$\dag$}\footnotetext{This work was conducted during his internship at Houmo AI.}

\begin{abstract}
Large language models (LLMs), with their billions of parameters, pose substantial challenges for deployment on edge devices, straining both memory capacity and computational resources. 
Block Floating Point (BFP) quantisation reduces memory and computational overhead by converting high-overhead floating point operations into low-bit fixed point operations.
However, BFP requires aligning all data to the maximum exponent, which causes loss of small and moderate values, resulting in quantisation error and degradation in the accuracy of LLMs.
To address this issue, we propose a Bidirectional Block Floating Point (BBFP) data format, which reduces the probability of selecting the maximum as shared exponent, thereby reducing quantisation error.
By utilizing the features in BBFP, we present a full-stack \underline{B}idirectional \underline{B}lock Floating Point-Based Quantisation \underline{A}ccelerator for \underline{L}LMs (BBAL), primarily comprising a processing element array based on BBFP, paired with proposed cost-effective nonlinear computation unit.
Experimental results show BBAL achieves a 22\% improvement in accuracy compared to an outlier-aware accelerator at similar efficiency, and a 40\% efficiency improvement over a BFP-based accelerator at similar accuracy.

\end{abstract}

\section{Introduction}
\label{sec:Intro}

LLMs\cite{LLMs} have achieved remarkable success, including text generation\cite{textgenerate}, text understanding\cite{textunderstanding}, and language translation\cite{translate}. 
However, the substantial size and high computational cost of these models constrain their deployment on edge devices.
For instance, deploying a Llama-70B \cite{touvron2023llama}  model necessitates the use of at least two 80GB A100 GPUs. With the continuous growth in the size of models, deploying them on edge devices imposes higher overhead.
Thus, reducing computational cost and storage demands has emerged as a critical challenge that needs to be addressed\cite{llminfe,llmhardware}.

\begin{figure}[t]
\centering

\includegraphics[width=0.485\textwidth]{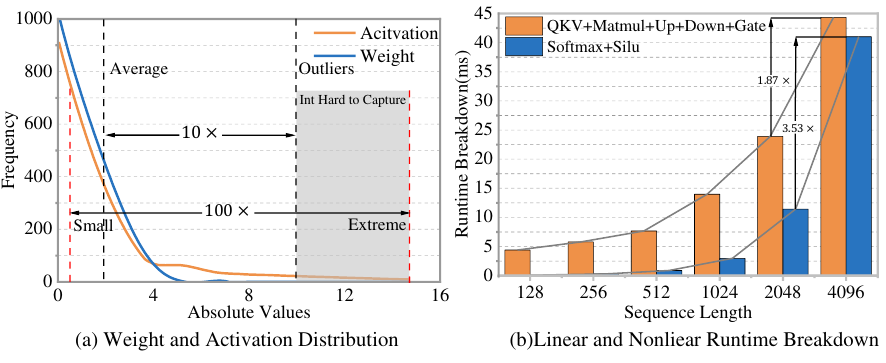}

\caption{(a) Distribution of activation and weight values in OPT-6.7B. (b) Linear and nonlinear runtime in the decoder stage of Llama-7B.}

\label{fig:fig1}

\end{figure}

Various techniques have been investigated \cite{deepcompress} to facilitate the efficient deployment of LLMs.
Among them, quantisation\cite{huang2024billm,dettmers2022gpt3,yuan2023rptq} is one of the most effective methods for reducing inference cost.
Especially, INT quantisation has been proven to significantly reduce storage requirements and computational cost when compared to the floating point. 
Yet, due to the limitations imposed by the representational range of INT and the presence of outliers in LLMs (see Fig.~\ref{fig:fig1}~(a)), INT quantisation usually faces significant accuracy degradation.
For instance, INT8 \cite{achiam2023gpt} quantisation increases perplexity (PPL) over 20\%.

To address this issue, researchers have proposed several data formats with a wider dynamic range to capture the outliers;
examples include BF16 \cite{bfloat16}, FP8 \cite{fp8}, etc.
Among these, Block Floating Point (BFP) stands out as a promising quantisation method \cite{rethinkbfp8,afp,blockminifloat,bie}, offering a favourable trade-off between performance and hardware overhead.
By compelling a set of fixed point numbers to share a common exponent, BFP effectively converts floating point operations into fixed point computations. 
Current research on BFP primarily focuses on accelerating computations in linear layers\cite{bfpcnn,bfpcnn2,bfperror}, and the nonlinear layers are entirely ignored.
However, Transformer-based models\cite{attentionisallyouneed}, contain numerous nonlinear operations, including Softmax and SILU, which often involve transcendental function computations and exhibit lower robustness, typically necessitating a broader representation range, high-precision, and higher-overhead floating point computations.

As revealed in Fig.~\ref{fig:fig1}~(b), the computation time taken by nonlinear operators increases with the growth of input sequence length, gradually becoming a performance bottleneck\cite{stevens2021softermax}. 
Since BFP offers a similar representation range to floating point, with greater computational efficiency, it holds the potential for simplifying operations in nonlinear layers. 
However, BFP's strategy of aligning all data to the maximum exponent introduces moderate quantisation error. 
This limits its ability to achieve lower-bit BFP quantisation in linear layers and hinders its application in nonlinear computations. 
For instance, directly applying BFP4 \cite{rethinkbfp8} quantisation to the linear layers of the OPT\cite{zhang2022opt} model can result in over a 40\% increase in PPL. Similarly, as shown in Table~\ref{nonlinear eval}, using BFP for computations in nonlinear layers can lead to a $5\times$ increase in PPL.

\vspace{-2pt}
\noindent \textbf{Contributions:} To reduce the quantisation error caused by BFP's alignment strategy and explore its potential applications in nonlinear layers, we propose a block-based data format, Bidirectional Block Floating Point (BBFP). 
By employing a 1-bit flag to distinguish between high and low mantissas, not all data needs to be aligned to the maximum exponents, reducing the probability of selecting a larger exponent as the shared exponent. This enables BBFP to capture outliers while minimizing quantisation error for moderate and small values.
Finally, based on BBFP, we propose BBAL, primarily comprising an optimized PE array based on BBFP, paired with our proposed cost-effective nonlinear computation unit. Experimental results show that BBAL achieves a 22\% improvement in accuracy compared to an outlier-aware accelerator at similar efficiency and a 40\% efficiency improvement over a vanilla BFP-based accelerator at similar accuracy. The main contributions presented are as follows:

\begin{itemize}

 \item \textbf{Data Format:} We propose BBFP to reduce BFP quantisation error. We also explore the impact of different shared exponent selections and overlap bit configurations on the quantisation error of BBFP. Finally, based on the characteristics of BBFP, we design an efficient basic computation unit.

  \item \textbf{Nonlinear Unit:} Leveraging the low quantisation error and fixed point computation characteristics of BBFP, we propose an efficient nonlinear computation unit based on BBFP. Additionally, benefiting from the shared exponent feature of BBFP, we introduce an exponent-based segmented lookup table method, which improves compatibility and reduces consumption. 
 
  \item \textbf{LLMs Accelerator:} 
Finally, we present BBAL, primarily comprising an optimized PE array based on BBFP for efficient linear computation, paired with our proposed cost-effective nonlinear computation unit to reduce resource consumption and latency.

\end{itemize}

\section{Background}
\label{sc:Back}

\subsection{Quantisation}

Currently, quantisation data formats focus on lower-bit formats, such as floating point FP4\cite{fp4}, and fixed point like INT4\cite{illm}, both of which provide high memory efficiency. However, due to the lower arithmetic density of floating point, researchers prefer using fixed point formats for quantisation. Methods like SmoothQuant\cite{xiao2023smoothquant} and GPTQ\cite{frantar2022gptq} have effectively utilized fixed point for quantisation.
Nonetheless, the inherent limitations of integer representations and the sensitivity to outliers present significant challenges in achieving high performance with lower-bit fixed point quantisation.

\begin{figure}[t]

\centering
\includegraphics[width=0.48\textwidth]{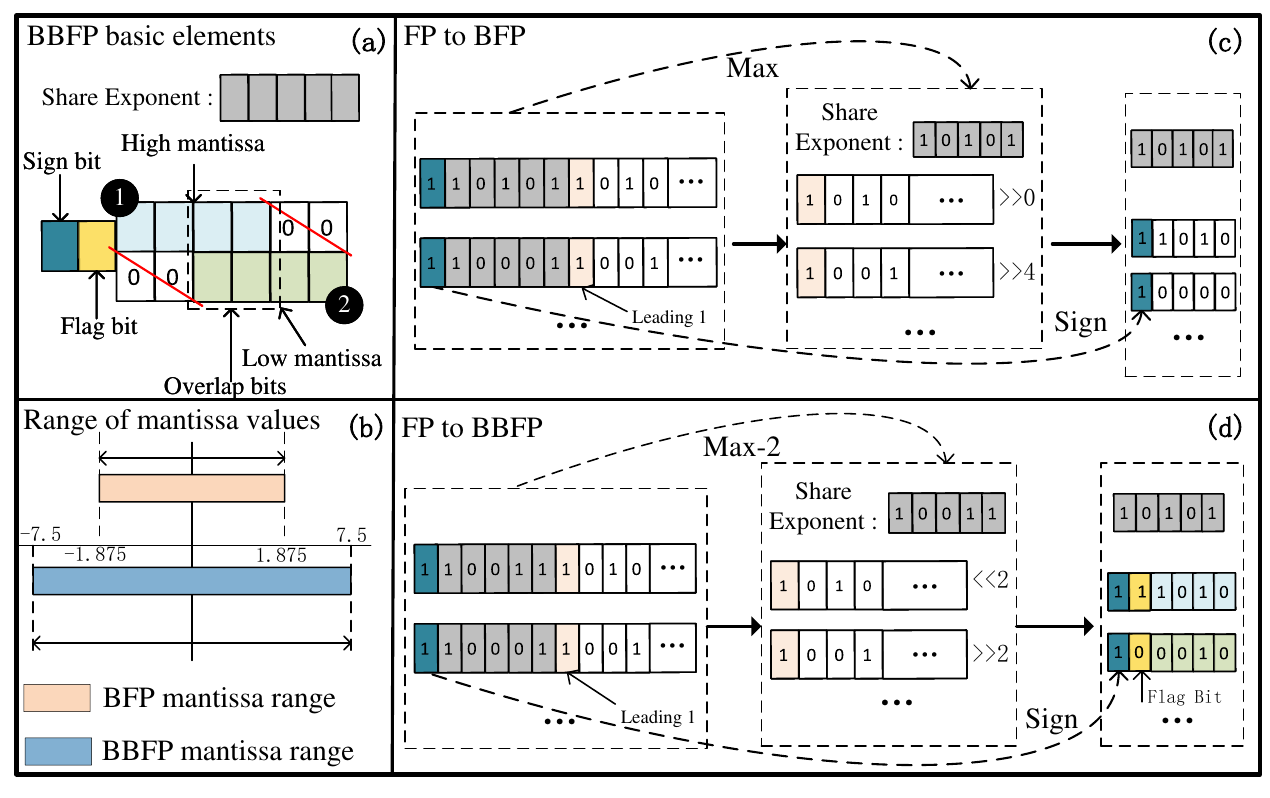}
\caption{(a) The basic components of BBFP(4,2); (b) comparing the representational range of mantissas between BBFP and BFP;
(c) the FP to BFP process; (d) the FP to BBFP process.}

\label{fig:BBFP}
\vspace{-4pt}
\end{figure}
Quantisation methods can generally be categorized into two approaches: Post-Training Quantisation (PTQ)\cite{liu2021post} and Quantisation-Aware Training (QAT)\cite{zafrir2019q8bert}. While QAT can maintain better accuracy at the same data width, the substantial model size of LLMs results in significant training cost, making PTQ a concise and effective quantisation method. In this work, we adopt the PTQ and propose BBFP that allows weight-activation quantisation without calibration.

\subsection{Block Floating Point}

In the IEEE-754 standard, each single-precision floating point number is composed of three parts: a 1-bit sign ${s}$, an 8-bit exponent ${e}$, and a 23-bit mantissa ${m}$. The actual value represented by the floating point number can be summarized as: $v=(-1)^{s}\times1.m\times2^{e-e_{bias}}$. Thus, for a vector of elements, the floating point representation is:
\vspace{-4pt}
\begin{equation}
[(-1)^{s_0} 2^{e_0} m_0 , (-1)^{s_1} 2^{e_1} m_1 , ... ,(-1)^{s_{n-1}} 2^{e_{n-1}} m_{n-1}]
\end{equation}

An format for optimizing performance and improving memory density is Block Floating Point (BFP). 
As shown in Fig.~\ref{fig:BBFP}~(c), BFP within a block share a max exponent by shifting the mantissa, and can be expressed as follows:
\begin{equation}
\begin{matrix}2^{e_m}[(-1)^{s_0}m_0',(-1)^{s_1}m_1',...,(-1)^{s_{N-1}}m'_{N-1}]\end{matrix}
\end{equation}
Where $2^{e_m}$ is the maximum exponent within a block of data, and ${m_i'}$ is the shifted mantissa. Thus, the dot product of two vectors in BFP format can be expressed as follows:

\begin{equation}
2^{e_{m_1}+e_{m_2}}\sum_{i=0}^{N-1}((-1)^{s_{1,i}\oplus s_{2,i}}m_{1,i}\cdot m_{2,i})
\end{equation}
where ${e_{m_1}}$ and ${e_{m_2}}$ are the shared exponent of two BFP vectors, ${\oplus}$ is an XOR operation. 

By transforming the complex floating point dot product operation into a fixed point equivalent, BFP significantly enhances computational efficiency. However, this improvement comes at the cost of reduced precision for small and moderate values, leading to a modest reduction in accuracy when quantizing LLMs.

\section{The Proposed Data Format}
\label{sc:bbfp}

\subsection{BBFP Data Format}

\label{3a}

\begin{figure}[t]
\centering
 
\includegraphics[width=0.49\textwidth]{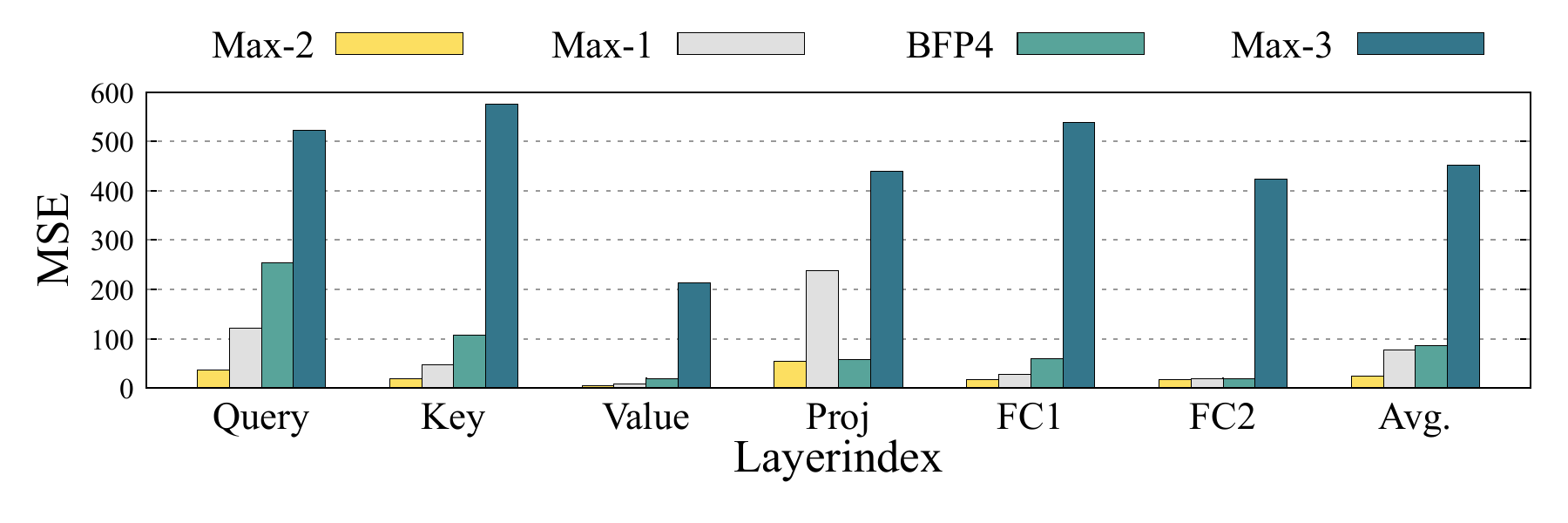}

\caption{Comparison of the impact of different selection of shared exponent of activation quantisation error with BBFP
(4,2).}

\label{fig:mse}
\vspace{-4pt}
\end{figure}

\noindent\textbf{BBFP Definition:} To reduce precision loss in small and moderate  values during BFP computations, we propose a data format named BBFP. Fig.~\ref{fig:BBFP}~(a) illustrates the proposed BBFP data format, which consists of a sign bit, a flag bit, ${e}$ bits for the shared exponent, ${m}$ bits for the mantissa, and ${o}$ bits for the overlap.
Throughout this paper, we denote different configurations of BBFP as BBFP(${m}$,${o}$) and different mantissa bit-width of BFP as BFP${m}$, where ${m,o\in\mathbb{N}}$. In all configurations, the shared exponent bit-width is fixed at 5 bits. 

In BBFP, the 1-bit flag serves to indicate whether the mantissa should be left- or right-shifted during alignment, while the ${o}$ bits of overlap are used to reduce error introduced by truncation when shifting left. 
The FP16 with an 11-bit mantissa and implicit leading one to BBFP(4,2) conversion can be summarized as follows:
\begin{equation}
    x_{\text{BBFP(4,2)}}=\begin{cases}Clip(x<<n)_{13,10}\,,\text{Flag}=1\\Clip(x>>n)_{11,8}\,\,\,,\text{Flag}=0\end{cases}
    \label{fp2bbfp}
\end{equation}
where {${Clip(\cdot)_{13,10}}$} and ${Clip(\cdot)_{11,8}}$ represent truncating the original mantissa from bit 13 to bit 10 and from bit 11 to bit 8. ${n}$ represents shift count and Flag represents a 1-bit flag in BBFP. 
 Due to the presence of the 2-bit overlap, truncation does not begin from the 12th bit to distinguish between left- (high) and right-shifted (low) mantissas, but rather from the 10th bit.
Regardless of whether the mantissa is shifted left or right, it is truncated to 4 bits and stored in memory. 
For example, when converting FP16 to BBFP without overlap bits, the most significant bit of the original mantissa is stored in the high-mantissa group, while the remaining bits are truncated and discarded. However, with the addition of two overlap bits, truncation starts from the 10th bit of the original mantissa, preserving 3 bits of information and thereby reducing quantisation error.




Fig.~\ref{fig:BBFP}~(d) shows the process of converting FP to BBFP in detail, which is similar to FP to BFP shown in Fig.~\ref{fig:BBFP}~(c). 
First, a shared exponent is determined, using the ${Max-2}$ exponent as an example. Next, the origin exponent is compared with the shared exponent and the origin mantissa 
is shifted. If the original floating point exponent is greater than the shared exponent, the flag is set to 1, and the mantissa is left-shifted. When the floating point exponent is less than or equal to the shared exponent, the flag is set to 0, and the mantissa is either right-shifted or remains unchanged. Finally, the shifted mantissa is truncated to four bits. Thus, the value represented by BBFP can be summarized by the following equation:
\begin{equation}
\begin{matrix}2^{e_s}[(-1)^{s_0}m_0'\times f,(-1)^{s_1}m_1'\times f,...,(-1)^{s_{N-1}}m'_{N-1}\times f]\end{matrix}
\end{equation}

\begin{equation}
f=\begin{cases}1,&\text{Flag}=0\\2^{m-o},&\text{Flag}=1\end{cases}
\end{equation}
Where ${e_s}$ represents the shared exponent, ${m_i'}$ represents the shifted mantissa, Flag indicates the flag bit, ${m}$ represents the bit width of the mantissa, and ${o}$ represents the bit width of the overlap.
From the above equation, it is evident that, given the same mantissa width, BBFP provides enhanced representational capability for the mantissa compared to BFP, as illustrated in Fig.~\ref{fig:BBFP}~(b).

\noindent\textbf{BBFP Dot Product:} The dot product of two vectors in BBFP format can be expressed as follows:

\begin{equation}
2^{e_{s_1}+e_{s_2}}\sum_{i=0}^{N-1}((-1)^{s_{1,i}\oplus s_{2,i}}m_{1,i}{f_{1,i}}\cdot m_{2,i}{f_{2,i}})
\label{dotbbfp}
\end{equation}
the above expression demonstrates that BBFP retains the characteristic of BFP that converts floating-point operations into fixed-point operations. By incorporating multiplexer and shifting modules, the 
mantissa representational range increases by 4${\times}$.

\begin{figure}[t]
\centering

\includegraphics[width=0.485\textwidth]{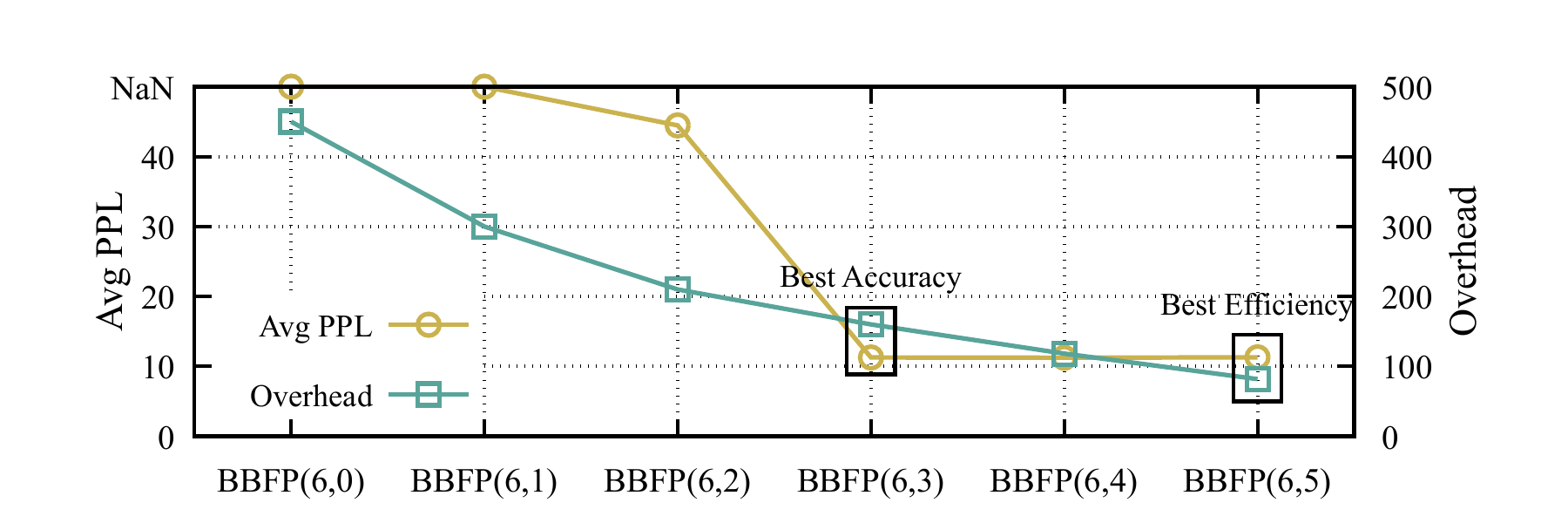}
\caption{The PPL and hardware overhead for BBFP with a width of 6 under varying overlap bit-widths.}

\label{fig:overlap}

\end{figure}

\subsection{Quantisation Error Analysis}
To compare the quantisation error with BFP, we analyse the sources of error in block floating point quantisation. For a block floating point using the round to nearest scheme, its quantisation error is zero-mean, and its variance ${\sigma^2}$ can be described as follows\cite{roundofferror}:

\begin{equation}
\sigma^2=\frac{2^{-2L_m}}{12}\sum_{i=1}^{N_\gamma}p_{\gamma_i}2^{2\gamma_i}
\label{bfperror}
\end{equation}

\begin{algorithm}[b]
\caption{Selection of overlap bit width}
\textbf{Input:} LLMs $Model$, Weight of overhead $w$,  Width of Mantissa $m$. \\
\textbf{Output:} Optimal Overlap Bits $o$.
\begin{algorithmic}[1]
    \State \textbf{Function} \texttt{Select\_Best\_Width}($Model$, $w$, $m$)
    \For{$i = 0$ to $m-1$}
        \State \hspace{1em} $PPL[i]=$Calculate\_PPL($model$,BBFP($m$,$i$))
        \State \hspace{1em} $Overhead[i]=$Calculate\_Overhead(BBFP($m$,$i$))
    \EndFor
    \For{$j = 0$ to $m-1$} \Comment{\#Max Norm and Cal score}
        \State \hspace{1em} $PPL[j]=PPL[j]/Max(PPL)$
        \State \hspace{1em} $Overhead[j]=Overhead[j]/Max(Overhead)$
        \State \hspace{1em} $score[j]=w \times Overhead[j]+(1-w)\times PPL[j]$
    \EndFor
    \State    $o=Min(score).index$
    \State \Return $o$
    \State \textbf{End Function}
\end{algorithmic}
\label{alg}
\end{algorithm}

Where ${L_m}$ denotes the length of the mantissa, and ${p_{\gamma_i}}$ represents the probabilities mass function of the block exponent. ${N_\gamma=2^{L_E}}$ is the number of available block exponent levels. When ${L_m}$ are the same for both BFP and BBFP, the only factor influencing the quantisation error is ${p_{\gamma_i}}$. Compared to the operation of aligning to the maximum exponent in BFP, BBFP allows for alignment to the non-maximum exponent, 
decreasing the quantisation error variance.

\subsection{Selection of Shared Exponent}
As described in Eq.~\eqref{fp2bbfp}, converting FP16 to BBFP involves shifting and truncating the mantissa. Intuitively, higher bits of data have more significance, so protecting these higher bits during the shift and truncation process can reduce quantisation error. 
However, as shown in Eq.~\eqref{bfperror}, reducing the probability of selecting the maximal shared exponent can minimize quantization error caused by right shift and truncation. Therefore, based on the two factors mentioned above, the selection of shared exponent we propose is as follows:
\begin{equation}
        E_{\text{shared}}=Max(E)-(m-o)
    \label{eq:alignment}
\end{equation}

\noindent where ${Max}$ denotes taking the maximum value, ${m}$ represents the width of the mantissa, ${o}$ and represents the width of the overlap bits. 
Fig.~\ref{fig:mse} compares the quantisation error of BBFP with a 4-bit mantissa and different shared exponent and BFP4. 
The $max-3$ alignment strategy, which is defined as $max-(m-o)-1$, results in significant error due to the left shift of the most significant bit, moving it out of the truncation range. By contrast, the $max-1$ alignment strategy, defined as $max-(m-o)+1$, is more likely to select larger values as the shared exponent compared to the $max-(m-o)$ alignment strategy, leading to more error. 

\subsection{Selection of Overlap Bit Width}

When the shared exponent selection strategy is set, increasing the width of overlap bit can reduce error from truncation due to left shift. However, according to Eq.~\eqref{eq:alignment} a wider overlap increases the probability of encountering the maximal shared exponent, which leads to greater loss for moderate and small values. Moreover, since various LLMs exhibit distinct data distribution and sensitivities to numerical error, and different overlap widths result in different hardware cost, we propose Algo.~\ref{alg} to determine the overlap widths.

By adjusting the overhead weight in Algo. 1, we can balance the priorities of accuracy and hardware cost. Fig.~\ref{fig:overlap} shows the process of optimizing the overlap width for BBFP with a width of 6. 

\begin{figure}[t]
\raggedright

\includegraphics[width=0.485\textwidth]{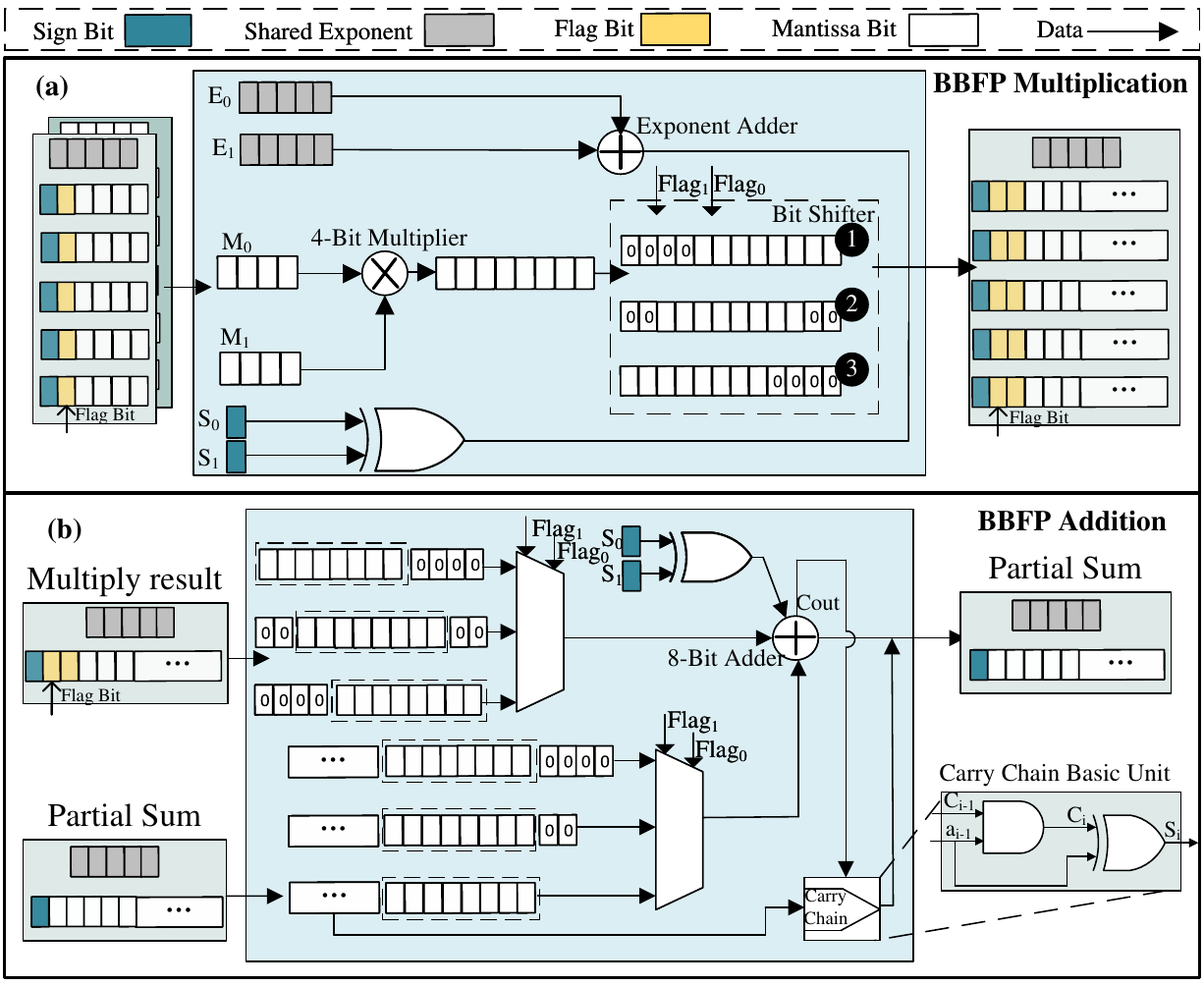}
\caption{(a) Multiplication operation between
two BBFP blocks; (b) partial sum operation with BBFP and carry chain module.}

\label{fig:basic}
\vspace{-5pt}
\end{figure}

\section{Data Format-Driven Hardware Design}

\subsection{Linear Computing Unit}

The Multiply and Accumulate (MAC) module is a fundamental computational unit in LLMs, primarily involving inter-block multiplication and partial-sum addition. Given that the proposed data format exhibits structured bit-level sparsity during computation, we optimize the MAC module to improve computational density.

\noindent\textbf{Intra-Block Multiplication:} As shown in Fig.~\ref{fig:basic}~(a), the inter-block matrix multiplication in BBFP can be divided into shared exponent addition and block mantissa multiplication. The mantissa multiplication can be expressed as follows:

\begin{equation}
m_{0}=\begin{cases}m_{1}\times m_{2}\qquad\quad,\text{Flag}_{1}\& \text{Flag}_{2}=0\\m_{1}\times m_{2}<<2\,\,,\text{Flag}_{1}\oplus \text{Flag}_{2}=1\\m_{1}\times m_{2}<<4\,\,,\text{Flag}_{1}\& \text{Flag}_{2}=1\end{cases}
\end{equation}

\noindent Where ${F_{1}}$,${F_{2}}$ represent the two Flag bits. In the case of BBFP(4,2), the multiplication of the two 4-bit mantissas can be performed using a 4-bit multiplier, followed by a shift operation to obtain a 12-bit mantissa, as shown in Fig.~\ref{fig:basic}~(a), where four bits are constant zero.
To improve memory density, these zero bits are removed, and a 2-bit flag is used to represent the zero elements. 
For example, as shown in Fig.~\ref{fig:basic}~(a), a flag of 00 corresponds to \circled{1} , while 01 or 10 corresponds to \circled{2} and 11 corresponds to \circled{3}. 
So, the final output is a BBFP consisting of a 2-bit flag, 1-bit sign, and 8-bit mantissa.

\noindent\textbf{Partial-Sum Addition:} In the MAC module, after performing inter-block multiplication, partial-sum addition is required. We observed that the resulting data blocks from inter-block multiplication exhibit a regular bit-level sparsity pattern. Therefore, we employ sparse addition to reduce the adder bit-width and decrease resource consumption. Fig.~\ref{fig:basic}~(b) illustrates three types of typical sparse adders. An 8-bit adder combined with a n-bit carry chain is used to replace a ${12+n}$-bit adder. The full adder expression is as follows:

\begin{equation}
    S=CI\oplus a_{i} \oplus b_{i}
    \label{eq:e1}
\end{equation}
\begin{equation}
    C = a_{i}b_{i}+C_{i}(a_{i} \oplus b_{i})
    \label{eq:e2}
\end{equation}

Where ${a_i}$ and ${b_i}$ represent the i-th bit of the partial sum and the multiplication result, respectively. As shown in Fig.~\ref{fig:basic}~(b), ${a_i}$ is not always zero, whereas ${b_i}$ becomes zero under specific patterns. This allows the adder to be simplified as follows:

\begin{equation}
    S_{c}=C_{i}\oplus a_{i}
    \label{eq:e3}
\end{equation}
\begin{equation}
    C_{co}=C_{ci}a_{i}
    \label{eq:e4}
\end{equation}
Comparing Eq.~\eqref{eq:e4} and Eq.~\eqref{eq:e3} with Eq.~\eqref{eq:e2} and Eq.~\eqref{eq:e1}, the carry chain module reduces one AND gate and two XOR gates compared to the full adder module. For example by replacing the 12-bit adder with an 8-bit adder and a 4-bit carry chain, the adder unit achieves a 15\% reduction in resource consumption. 
Furthermore, as the BBFP bit-width increases and the number of overlapping bits decreases, the optimization effect becomes increasingly pronounced.
\begin{table}[t]
    \centering
    \caption{Various data types for MAC unit Mem Eff. and area.}
    \begin{tabular}{>{\centering\arraybackslash}p{1cm}>{\centering\arraybackslash}p{0.8cm}>{\centering\arraybackslash}p{0.8cm}>{\centering\arraybackslash}p{2.6cm}>{\centering\arraybackslash}p{1.5cm}}
    \toprule
        \textbf{Datatype} & \textbf{BlockSize} & \textbf{Area} &\textbf{Equivalent Bit-Widt}h& \textbf{Mem Eff.} \\ 
    \midrule
        FP 16 & 1 & 39599 &16& 1${\times}$ \\ 
        INT 8 & 1 & 9257 &8& 2${\times}$ \\ 
        BFP 8 & 32 & 9371 &9.16& 1.75${\times}$ \\ 
        BFP 6 & 32 & 5633 & 7.16&2.24${\times}$ \\ 
         \rowcolor{lightgray}
        BBFP(8,4) & 32 & 9806 & 10.16&1.58${\times}$ \\ 
         \rowcolor{lightgray}
        BBFP(6,3) & 32 & 5764 & 8.16&1.96${\times}$ \\ 
    \bottomrule
    \end{tabular}
    \label{tab:mac}
\vspace{-5pt}
\end{table}

\noindent\textbf{MAC Efficiency:} Table.~\ref{tab:mac} presents the MAC unit area and memory efficiency across different formats. Compared to BFP, the area for BBFP increases due to the separation of high and low mantissa groups. This leads to wider widths for both multiplication and partial-sum addition, resulting in higher area consumption. 
Additionally, since BBFP introduces an extra 1-bit flag bit, its memory efficiency is slightly lower. 
However, BBFP(6,3) offers the higher representation capability than BFP8, while consuming less area and memory footprint. 
This demonstrates that our new format provides stronger representational power and lower computational overhead.

\subsection{Proposed nonlinear computing unit}

\noindent\textbf{Segment Lookup Table: } 

Unlike FP32, where each data value has an exponent field, BBFP assigns a shared exponent field for a set of data. 
Therefore, we propose a segmented approach to load LUT,  based on the shared exponent and perform lookups according to mantissas. 
Firstly, we divide the function values into several segments, based on different exponents, and store these segments in external memory. 
For instance, with five exponent bits, the function is split into $2^5\times 2$ sub-tables.
Secondly, once a shared exponent is calculated during the alignment phase, the corresponding sub-table can be loaded. 
Finally, unlike floating point LUT, which require additional mapping, BBFP uses the mantissa directly as the address for lookup.

\noindent\textbf{Pipelined Design}: To improve hardware throughput and mask the time required for loading LUT from external memory, the entire nonlinear computation unit is designed with a pipelined architecture. 
Each module is equipped with a buffer. 
Additionally, the computation unit is capable of computing different transcendental functions. 
However, different functions may require different computation sequences and computation units. 
To address this, the computation unit features adjustable computation order and is equipped with redundant units.

To illustrate the data flow adjustment for computing the sigmoid function, the formula for the sigmoid function is as follows:
\begin{equation}
    Sigmoid(x)=1/{(1+e^{-x})}
\end{equation}

First, the values of ${(1+e^{-x})}$ are computed offline, and the results are stored in external memory. The Control Unit then configures the data flow through the pipeline, directing it from the Align Exponent Unit to the LUT File, followed by the Div Unit, and finally to the Output Encoder. Similarly, the nonlinear computation unit can also compute functions such as SILU and GELU.
\begin{figure}[t]
\centering
\includegraphics[width=0.485\textwidth]{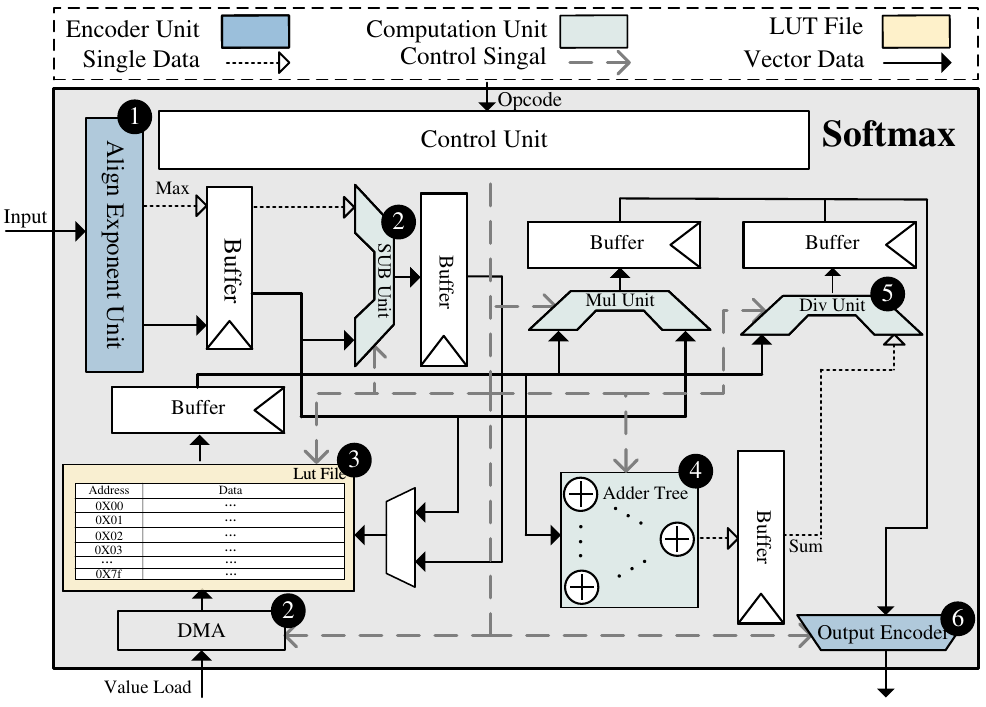}
\caption{Proposed nonlinear unit with Softmax as an example and the numbers in the figure indicate the computation sequence.
}
\label{fig:noliear}
\vspace{-5pt}
\end{figure}

\noindent\textbf{INT Computation:} After the data passes through the Align Exponent module, it is converted from FP16 to BBFP for lookup table computation. By  adjusting the data format in the table offline, each entry in the sub-table can be converted from FP16 to BBFP. This ensures that after the data passes through the LUT file, it retains the BBFP format for the next computation step.

\noindent\textbf{Basic Unit:} Based on the methods outlined above, we propose the nonlinear unit shown in Fig.~\ref{fig:noliear}. This nonlinear unit primarily consists of the Align Exponent Unit, Sub Unit, Mul Unit, Adder Tree, Div Unit, and LUT File. Fig.~\ref{fig:noliear} illustrates the flow of the nonlinear unit using the softmax computation as an example.

\subsection{LLMs Accelerator}

\begin{figure}[t]
\centering

\includegraphics[width=0.485\textwidth]{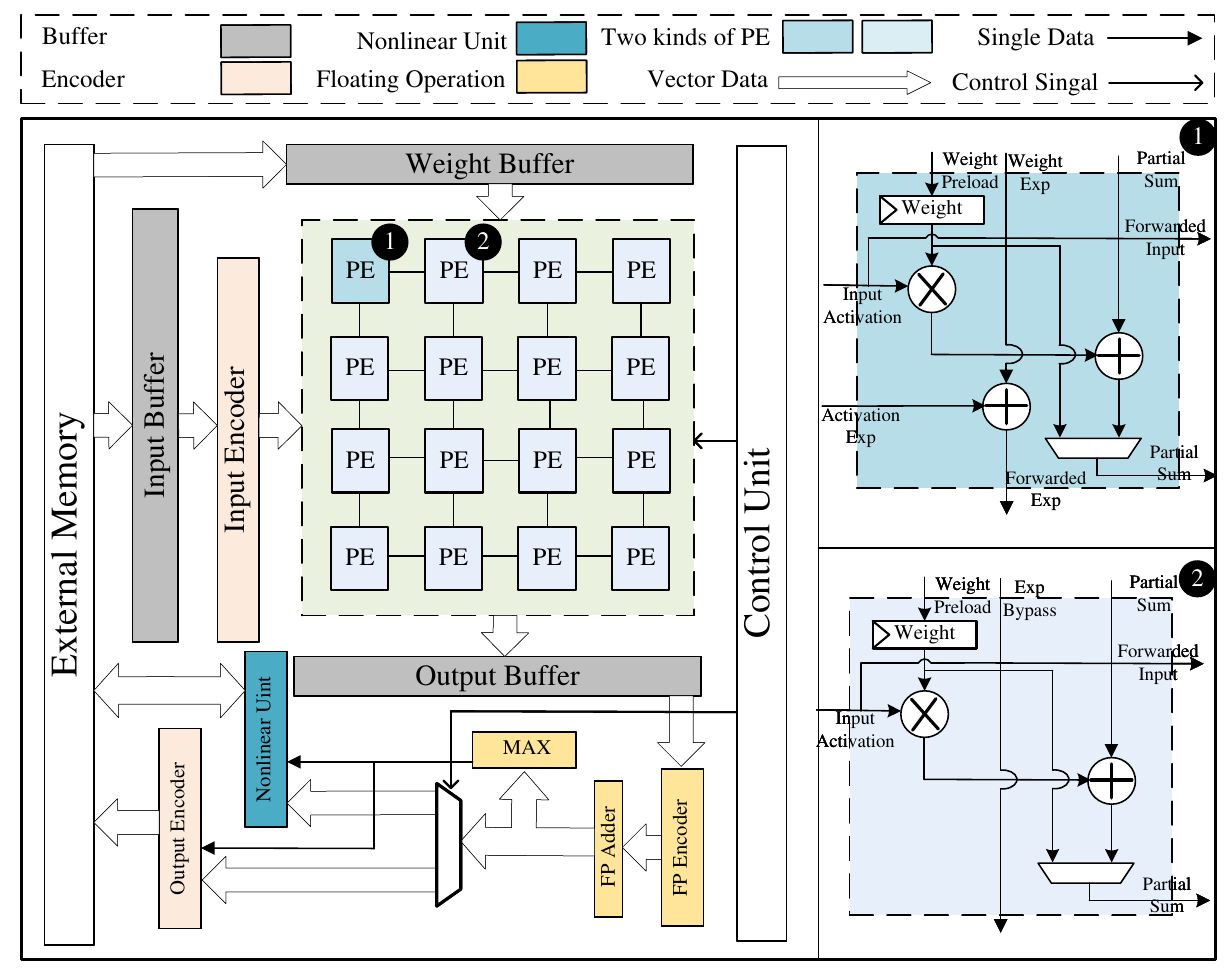}
\caption{ A computation accelerator based on BBFP, incorporating two
types of PE optimized for BBFP operations.}
\label{fig:acceleator}
\vspace{-5pt}
\end{figure}

\noindent \textbf{Overall Architecture:} The BBAL is shown in Fig.~\ref{fig:acceleator}, which is centred around a weight stationary PE array and proposed nonlinear computation unit, accompanied by an input encoder, input buffer, weight buffer, output buffer, output encoder, max unit, FP encoder, FP adder, and control unit. Performing multiplication between two BFP requires only a single exponent addition. Therefore, the PE computation units have been partially modified to accommodate block floating point operations, resulting in two versions: \circled{1} adds a shared exponent adder, and \circled{2} includes a shared exponent bypass unit.

\begin{table*}[t]
    \centering
    \caption{Perplexity results of quantized model on Wikitext2 (lower is better).}

      \resizebox{0.98\textwidth}{!}{
    \begin{tabular}{>{\centering\arraybackslash}p{1.1cm}|>{\centering\arraybackslash}p{1.15cm} >{\centering\arraybackslash}p{1.15cm} >{\centering\arraybackslash}p{1.15cm} >{\centering\arraybackslash}p{1.29cm} >{\centering\arraybackslash}p{1.29cm} >{\centering\arraybackslash}p{1.29cm} >{\centering\arraybackslash}p{1.15cm} >{\centering\arraybackslash}p{1.15cm} >{\centering\arraybackslash}p{1.15cm} >{\centering\arraybackslash}p{1.15cm} >{\centering\arraybackslash}p{1.15cm} >{\centering\arraybackslash}p{1.15cm} }
    \toprule
        Model & Llama-1B & Llama-3B & Llama-7B & Llama-13B & Llama-30B & Llama-65B & OPT-1.3B & OPT-2.7B & OPT-6.7B & OPT-13B & OPT-30B & OPT-66B  \\ 
    \midrule
        FP16 & 9.88 & 7.87 & 5.47 & 5.09 & 4.10 & 3.53 & 14.62 & 12.47 & 10.86 & 10.12 & 9.56 & 9.34  \\ 
    \midrule
        Oltron & N/A & N/A & 14.67 & 9.48 & 7.51 & 6.69 & N/A & N/A & 11.99 & 11.65 & 10.60 & 10.29  \\ 
        Olive & N/A & N/A & 144.78 & 42.24 & 36.55 & NaN & N/A & N/A & 107.15 & 416.57 & 334.7 & 4058.83  \\ 
        Omniquant & N/A & N/A & 11.26 & 10.87 & 10.33 & 9.17 & N/A & N/A & 12.24 & 11.65 & 10.6 & 10.29  \\ 
    \midrule
        BFP6(6m) & 10.06 & 7.95 & 5.61 & 5.13 & 4.12 & 3.61 & 15.57 & 12.5 & 10.91 & 10.22 & 9.62 & 9.48  \\ 
        BFP4(4m) & 13.45 & 9.44 & 5.83 & 5.72 & 5.05 & 4.12 & 27.21 & 18.98 & 12.24 & 11.56 & 10.50 & 10.10  \\ 
    \midrule
        BBFP(3,1) & 12.35 & 9.00 & 5.66 & 5.33 & 4.46 & 4.01 &  23.12& 15.29 & 14.07 & 10.85 & 10.45 & 10.27  \\ 
        BBFP(4,2) & 10.41 & 8.13 & 5.80 & 5.39 & 4.37 & 3.65 & 17.06 & 13.36 & 12.03 & 10.39 & 9.63 & 9.87  \\ 
        BBFP(4,3) & 10.65 & 8.20 & 5.80 & 5.20 & 4.26 & 3.69 & 17.52 & 13.89 & 11.54 & 10.38 & 9.61 & 9.93  \\ 
         \rowcolor{lightgray}BBFP(6,3) & \textbf{9.93} & \textbf{7.89} & \textbf{5.48} & \textbf{5.09} & \textbf{4.10 }& \textbf{3.59} & \textbf{15.16} & \textbf{12.49} & \textbf{10.89} & \textbf{10.12} & \textbf{9.55} & 9.38  \\ 
        BBFP(6,4) & 9.93 & 7.9 & 5.48 & 5.09 & 4.10 & 3.59 & 15.00 & 12.47 & 10.89 & 10.14 & 9.55 &\textbf{ 9.36 } \\ 
    \bottomrule
    \end{tabular}
    }\label{tab:all ppl}
\end{table*}

\begin{table*}[t]
    \centering

    \caption{Comparison of PE area ($\mu m^2$) across different quantization strategies (normalized by maximum BBFP(6,3) PE area).}
   
    \resizebox{0.98\textwidth}{!}{
    \begin{tabular}{c|ccccccccccc}
    \toprule
        ~ & Oltron & Olive & BFP4 & BFP6 & BBFP(3,1) & BBFP(3,2) & BBFP(4,2) & BBFP(4,3) & BBFP(6,3) & BBFP(6,4) & BBFP(6,5) \\ 
    \midrule
        Area & 78.50 & 156.47 & 215.23 & 110.24 & 77.69 & \textbf{75.51} & 117.11 & 113.31 & 241.01 & 231.14 & 224.70 \\ 
        Norm & 0.33 & 0.65 & 0.46 & 0.90 & 0.32 & 0.31 & 0.49 & 0.47 & 1.00 & 0.96 & 0.93 \\ 
    \bottomrule
    \end{tabular}}
    
    \label{tab:pearea}
\end{table*}

\begin{table}[t]

    \centering
    \caption{PPL of Llama at various schemes with nonliear units.}
    
    \resizebox{0.48\textwidth}{!}{
    \begin{tabular}{l cccc}
    \toprule
        Data Format & nonlinear Operation & LLama-7B & Llama2-7B & Llama3-8B \\
        \midrule
        FP32 &Altogether&5.68&5.47&6.14\\
    \midrule
     \multirow{3}{*}{BBFP(10,5)} & Softmax Only & 5.74 & 5.62 & 6.24 \\
        & SILU Only & 5.71 & 5.53 & 6.21 \\
        & Altogether & 5.81 & 5.91 & 6.34 \\
    \midrule
        \multirow{3}{*}{BFP10} & Softmax Only & 67.31 & 32.72 & 69.95 \\
        & SILU Only & 33.21 & 17.54 & 31.30 \\
        & Altogether & 99.28 & 50.21 & 102.35 \\
    \bottomrule
    \end{tabular}
    }
    \label{nonlinear eval}
    
\end{table}

\begin{table}[t]
    \centering
    \caption{Comparison of ADP, EDP, Eff., and Compatibility.}
    
    \resizebox{\columnwidth}{!}{ 
    \begin{tabular}{>{\centering\arraybackslash}p{0.7cm}
    >{\centering\arraybackslash}p{0.7cm}
    >{\centering\arraybackslash}p{1.5cm}
    >{\centering\arraybackslash}p{1cm}
    >{\centering\arraybackslash}p{1.1cm}
    >{\centering\arraybackslash}p{1.1cm}
    >{\centering\arraybackslash}p{1.9cm}}
    \toprule
        \textbf{Methods} & \textbf{Num} & \textbf{Format} & \textbf{ADP↓} & \textbf{EDP↓} & \textbf{Eff.↑} & \textbf{Compatibility} \\ 
    \midrule
        \cite{Design1}& 10 & Int8 & ${\sim}$4.33 & ${\sim}$79.58 & ${\sim}$85.98 & - \\ 
        \cite{design2} & 8 & Int 27 & ${\sim}$299.13 & ${\sim}$18691.24 & 3.31 & - \\ 
        \rowcolor{lightgray} 
         Ours & 16 & BBFP(10,5,5) & 32.64 & 1040.40 & 98.03 & SILU and so on \\ 
    \bottomrule
    \end{tabular}
    }
    \label{tab:comparison}
     
\end{table}

\noindent \textbf{Computation Flow:} When processing matrix multiplication, each 4$\times$4 elements are encoded into BBFP and is sent to the PE array for computation. After computation, the data flows through the FP encoder, where it is encoded into FP format and sent to the FP adder, waiting to perform addition. After executing the addition, the data flows through a max unit and a data selector. The value output by the max unit can be used by either the nonlinear unit or the output encoder, eliminating the need for an additional comparator. Finally, the control unit determines whether the output from the floating point adder will pass through the nonlinear computation unit.

\section{Evaluation}

\subsection{Configuration}

\noindent\textbf{Nonlinear Configuration:} We used BBFP(10,5) to quantise the nonlinear layers, with the address width of each LUT being 7-bit. 
Considering the tendency of Softmax values towards zero, we designed 18 sub-tables, with SILU having 24 sub-tables. 
We replaced the nonlinear layers in the Llama-7B\cite{touvron2023llama}, Llama2-7B\cite{touvron2023llama2}, and Llama3-8B\cite{meta2024llama3} and evaluated their performance on the WikiText2\cite{wiki}. 
To validate the accuracy of our nonlinear quantisation strategy, we used FP32-based nonlinear units as the baseline and compared them with BBFP(10,5) and BFP10. 
Furthermore, to assess the efficiency of our quantisation strategy, we compared it with sota methods from three perspectives: EDP, ADP, and Efficiency (Throughput/(Area$\times$Power)).

\noindent\textbf{Linear Configuration:} We modified the BBFP implementations to evaluate their impact on LLMs. 
The dataset used in this experiment was WikiText2.
To validate the accuracy of the quantisation methods, we used FP16 as the baseline and compared BBFP with mantissa widths of 3, 4, and 6 bits against BFP4, BFP6, and sota methods, including OmniQuant \cite{shao2023omniquant}, Oltron \cite{xue2024oltron}, and Olive \cite{guo2023olive}, all of which quantise both weights and activations. To assess the efficiency of the quantisation strategy, we made comparisons with other approaches across three dimensions: area, energy, and throughput.

\noindent\textbf{Implementation:} We implemented the design with BBFP and BFP using Chisel \cite{chisel} and evaluated it under the TSMC 28nm process. We utilized Design Compiler \cite{kurup1997logic}, and used CACTI \cite{muralimanohar2009cacti} to estimate the area and power of on-chip memories. We developed a cycle-level simulator
based on DnnWeaver\cite{sharma2016dnnweaver} to estimate the performance.

\subsection{Results Analysis:}

\noindent\textbf{Nonlinear Accuracy Analysis:} Table \ref{nonlinear eval} presents a comparison of PPL between BBFP(10,5) and BFP10. Experimental results demonstrate that our nonlinear quantisation strategy incurs a maximum PPL increase of only 0.44 across the three models, whereas BFP10 results in at least 3$\times$ PPL increases. This demonstrates that BBFP(10,5) bridges the gap between BFP and application to nonlinear operations.

\noindent\textbf{Nonlinear Efficiency Analysis:} Table~\ref{tab:comparison} compares our proposed nonlinear unit with other sota. Compared to low-precision approximation algorithms \cite{Design1}, our design shows a less favourable performance across the ADP and EDP.
This is because our proposed nonlinear computation unit for LLMs requires full-precision, high-bitwidth integer multipliers and dividers to minimize numerical error, which increases both area and power consumption. Additionally, to enhance compatibility, several redundant units are included; for example, the vector multiplication module remains idle during softmax computation, further contributing to larger area requirements and increased static power consumption.
However, leveraging a segmented-exponent dynamic lookup strategy allows the design to reduce costly on-chip memory by utilizing more affordable off-chip memory, achieving high compatibility and efficiency. Additionally, because BBFP  preserves the computational efficiency of fixed point, it results in significant conservation of computational resources. 
Experimental results demonstrate that our design achieves nearly a 30× efficiency improvement over high-precision method \cite{design2}.

\noindent\textbf{Linear Area Analysis:} Table~\ref{tab:pearea} presents the area of a PE under various methods. The PE area consists of two components: multiplier and adder, with multiplier occupying the majority. 
Hence, BFP, Olive, Oltron and BBFP, with the same mantissa width, exhibit similar areas.

\noindent\textbf{Linear Accuracy-Throughput Analysis:}
Table~\ref{tab:all ppl} shows the effects of BBFP quantisation for the linear layers without any calibration. 
Experimental results show that BBFP offers improved accuracy compared to BFP. 
Specifically, BBFP(3,1) achieves a 6\% improvement over BFP4, and BBFP(4,2) achieves an average PPL only 4\% higher than BFP6.
Additionally, BBFP(4,2) achieve 30\% lower PPL, compared to Oltron, and 33\%, compared to OmniQuant.

\begin{figure}[t]
\centering

\includegraphics[width=0.485\textwidth,height=3cm]{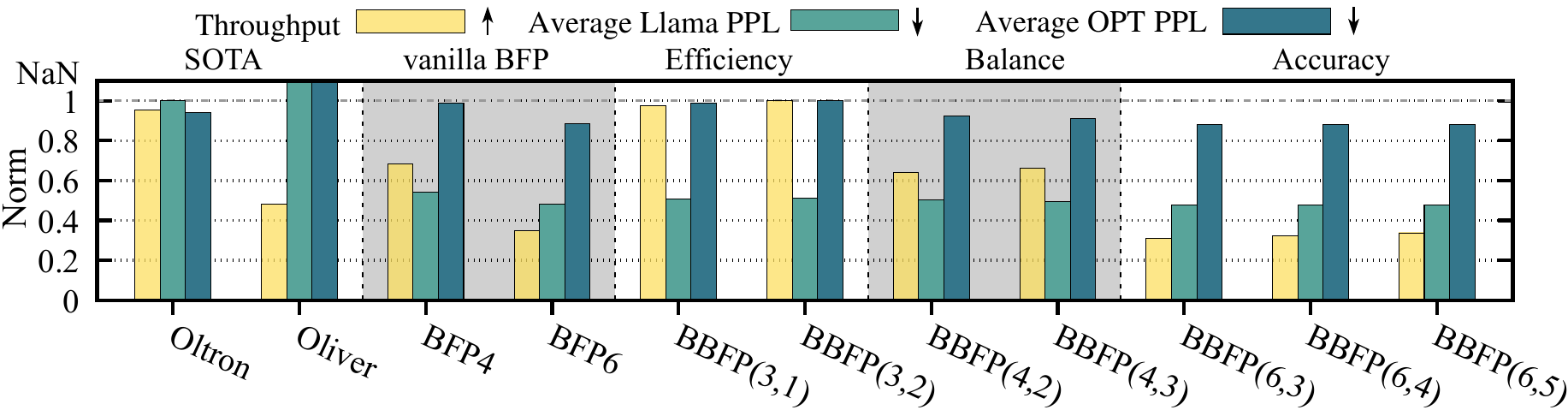}

\caption{Various quantisation methods compared by average Llama/OPT PPL (lower is better) and throughput for equal PE area.}
\label{fig:last}

\end{figure}

Through analyzing activation distribution across different models, we observe that models contain varying proportions and magnitudes of outliers. 
Hence, outlier-aware quantisation methods, which capture a fixed proportion of outliers, perform poorly on the Llama (with more outliers) but achieve better results on the OPT (with fewer outliers), as shown in Fig.~\ref{fig:last}.
BBFP's ability to capture any proportion of outliers ensures a stable accuracy baseline across models.

Fig.~\ref{fig:last} presents the accuracy and throughput performance of various quantisation strategies under the same PE iso-area condition. 
Since BBFP(3,1), BBFP(3,2), and Oltron all use 3-bit multipliers and low-bit adders, they exhibit similar throughput within the same area. However, due to BBFP’s superior outlier protection capability, BBFP(3,1) achieves a 22\% accuracy improvement on average across all tasks compared to Oltron. 
Additionally, compared to BFP4, BBFP(3,1) and BBFP(3,2) achieve a 40\% throughput improvement while maintaining similar accuracy. 

The BBFP with a width of 4 shows a 30\% drop in throughput compared to Oltron, but its PPL is reduced by 30\%.

\begin{figure}[t]
\centering

\includegraphics[width=0.485\textwidth,height=3cm]{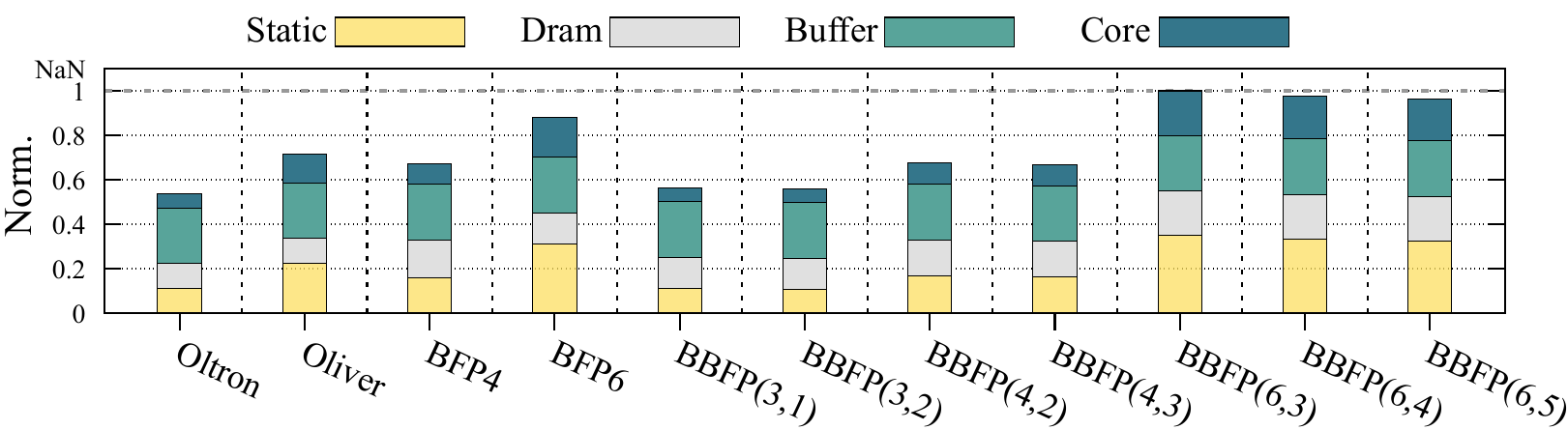}

\caption{
Average normalized energy consumption under identical PE count and buffer size (excluding nonlinear unit).}
\label{fig:power}

\end{figure}

\noindent\textbf{Energy Analysis:} Fig.~\ref{fig:power} illustrates the average energy consumption for various methods when executing tasks under the same number of PE and with the same buffer size.  Compared to BFP4, BBFP, with a width of 3, reduces energy consumption by 13\%, primarily attributed to reductions in both static and core energy.
The energy consumption increase of BBFP compared to BFP with the same mantissa bit-width is within 5\%. This increase is primarily due to the larger area, which results in higher static and dynamic power, as well as the additional 1-bit flag in BBFP contributing to slightly higher DRAM power.
\section{Conclusion}

In this paper, we propose BBFP, which is optimized based on BFP, to reduce quantisation error and promote the further application of BFP. The key insight is to reduce the probability of aligning maximum exponent as the shared exponent, thereby capturing outliers while minimizing quantisation error. 
Due to the low quantisation error of BBFP, we propose an efficient nonlinear computation unit based on BBFP, and this extends the application potential of BFP. Experimental results demonstrate that proposed nonlinear computation unit achieves a 30× efficiency improvement over high-precision nonlinear computation unit, with almost no accuracy loss.
Finally, we propose BBAL, primarily comprising an optimized PE array based on BBFP, paired with proposed nonlinear computation unit.
Experimental results demonstrate that BBAL achieves a 22\% accuracy improvement compared to outlier-aware accelerators with similar hardware consumption, and a 40\% throughput improvement compared to BFP4 with similar accuracy.

\section{Acknowledgement}
We appreciate the reviewers for their helpful feedback.
This work is supported by the National Key Research and Development Program (Grant No. 2024YFB4405600), the National Natural
Science Foundation of China (Grant No. 62472086), the Basic Research Program of Jiangsu (Grant No. BK20243042), the Science and Technology Major Special Program of Jiangsu (No. BG2024010), and the Start-up Research Fund of Southeast University (Grant No. RF1028624005).

\clearpage
\bibliographystyle{IEEEtran}
\bibliography{ref}

\begin{thebibliography}{10}
\providecommand{\url}[1]{#1}
\csname url@samestyle\endcsname
\providecommand{\newblock}{\relax}
\providecommand{\bibinfo}[2]{#2}
\providecommand{\BIBentrySTDinterwordspacing}{\spaceskip=0pt\relax}
\providecommand{\BIBentryALTinterwordstretchfactor}{4}
\providecommand{\BIBentryALTinterwordspacing}{\spaceskip=\fontdimen2\font plus
\BIBentryALTinterwordstretchfactor\fontdimen3\font minus \fontdimen4\font\relax}
\providecommand{\BIBforeignlanguage}[2]{{%
\expandafter\ifx\csname l@#1\endcsname\relax
\typeout{** WARNING: IEEEtran.bst: No hyphenation pattern has been}%
\typeout{** loaded for the language `#1'. Using the pattern for}%
\typeout{** the default language instead.}%
\else
\language=\csname l@#1\endcsname
\fi
#2}}
\providecommand{\BIBdecl}{\relax}
\BIBdecl

\bibitem{LLMs}
B.~Mann, N.~Ryder, M.~Subbiah, J.~Kaplan, P.~Dhariwal, A.~Neelakantan, P.~Shyam, G.~Sastry, A.~Askell, S.~Agarwal \emph{et~al.}, ``Language models are few-shot learners,'' \emph{arXiv preprint arXiv:2005.14165}, vol.~1, 2020.

\bibitem{textgenerate}
C.~Saharia, W.~Chan, S.~Saxena, L.~Li, J.~Whang, E.~L. Denton, K.~Ghasemipour, R.~Gontijo~Lopes, B.~Karagol~Ayan, T.~Salimans \emph{et~al.}, ``Photorealistic text-to-image diffusion models with deep language understanding,'' \emph{Advances in neural information processing systems}, vol.~35, pp. 36\,479--36\,494, 2022.

\bibitem{textunderstanding}
J.~Kil, S.~Changpinyo, X.~Chen, H.~Hu, S.~Goodman, W.-L. Chao, and R.~Soricut, ``Prestu: Pre-training for scene-text understanding,'' in \emph{Proceedings of the IEEE/CVF International Conference on Computer Vision}, 2023, pp. 15\,270--15\,280.

\bibitem{translate}
C.~Lyu, J.~Xu, and L.~Wang, ``New trends in machine translation using large language models: Case examples with chatgpt,'' \emph{arXiv preprint arXiv:2305.01181}, 2023.

\bibitem{touvron2023llama}
H.~Touvron, T.~Lavril, G.~Izacard, X.~Martinet, M.-A. Lachaux, T.~Lacroix, B.~Rozi{\`e}re, N.~Goyal, E.~Hambro, F.~Azhar \emph{et~al.}, ``Llama: Open and efficient foundation language models,'' \emph{arXiv preprint arXiv:2302.13971}, 2023.

\bibitem{llminfe}
Z.~Zhou, X.~Ning, K.~Hong, T.~Fu, J.~Xu, S.~Li, Y.~Lou, L.~Wang, Z.~Yuan, X.~Li \emph{et~al.}, ``A survey on efficient inference for large language models,'' \emph{arXiv preprint arXiv:2404.14294}, 2024.

\bibitem{llmhardware}
N.~Koilia and C.~Kachris, ``Hardware acceleration of llms: A comprehensive survey and comparison,'' \emph{arXiv preprint arXiv:2409.03384}, 2024.

\bibitem{deepcompress}
S.~Han, H.~Mao, and W.~J. Dally, ``Deep compression: Compressing deep neural networks with pruning, trained quantization and huffman coding,'' \emph{arXiv preprint arXiv:1510.00149}, 2015.

\bibitem{huang2024billm}
W.~Huang, Y.~Liu, H.~Qin, Y.~Li, S.~Zhang, X.~Liu, M.~Magno, and X.~Qi, ``Billm: Pushing the limit of post-training quantization for llms,'' \emph{arXiv preprint arXiv:2402.04291}, 2024.

\bibitem{dettmers2022gpt3}
T.~Dettmers, M.~Lewis, Y.~Belkada, and L.~Zettlemoyer, ``Gpt3. int8 (): 8-bit matrix multiplication for transformers at scale,'' \emph{Advances in Neural Information Processing Systems}, vol.~35, pp. 30\,318--30\,332, 2022.

\bibitem{yuan2023rptq}
Z.~Yuan, L.~Niu, J.~Liu, W.~Liu, X.~Wang, Y.~Shang, G.~Sun, Q.~Wu, J.~Wu, and B.~Wu, ``Rptq: Reorder-based post-training quantization for large language models,'' \emph{arXiv preprint arXiv:2304.01089}, 2023.

\bibitem{achiam2023gpt}
J.~Achiam, S.~Adler, S.~Agarwal, L.~Ahmad, I.~Akkaya, F.~L. Aleman, D.~Almeida, J.~Altenschmidt, S.~Altman, S.~Anadkat \emph{et~al.}, ``Gpt-4 technical report,'' \emph{arXiv preprint arXiv:2303.08774}, 2023.

\bibitem{bfloat16}
N.~Burgess, J.~Milanovic, N.~Stephens, K.~Monachopoulos, and D.~Mansell, ``Bfloat16 processing for neural networks,'' in \emph{2019 IEEE 26th Symposium on Computer Arithmetic (ARITH)}.\hskip 1em plus 0.5em minus 0.4em\relax IEEE, 2019, pp. 88--91.

\bibitem{fp8}
P.~Micikevicius, D.~Stosic, N.~Burgess, M.~Cornea, P.~Dubey, R.~Grisenthwaite, S.~Ha, A.~Heinecke, P.~Judd, J.~Kamalu \emph{et~al.}, ``Fp8 formats for deep learning,'' \emph{arXiv preprint arXiv:2209.05433}, 2022.

\bibitem{rethinkbfp8}
C.~Zhang, J.~Cheng, I.~Shumailov, G.~A. Constantinides, and Y.~Zhao, ``Revisiting block-based quantisation: What is important for sub-8-bit llm inference?'' \emph{arXiv preprint arXiv:2310.05079}, 2023.

\bibitem{afp}
T.~Yeh, M.~Sterner, Z.~Lai, B.~Chuang, and A.~Ihler, ``Be like water: Adaptive floating point for machine learning,'' in \emph{International Conference on Machine Learning}.\hskip 1em plus 0.5em minus 0.4em\relax PMLR, 2022, pp. 25\,490--25\,500.

\bibitem{blockminifloat}
S.~Fox, S.~Rasoulinezhad, J.~Faraone, P.~Leong \emph{et~al.}, ``A block minifloat representation for training deep neural networks,'' in \emph{International Conference on Learning Representations}, 2020.

\bibitem{bie}
L.~Zou, W.~Zhao, S.~Yin, C.~Bai, Q.~Sun, and B.~Yu, ``Bie: Bi-exponent block floating-point for large language models quantization,'' in \emph{Forty-first International Conference on Machine Learning}, 2024.

\bibitem{bfpcnn}
M.~Cho and Y.~Kim, ``Fpga-based convolutional neural network accelerator with resource-optimized approximate multiply-accumulate unit,'' \emph{Electronics}, vol.~10, no.~22, p. 2859, 2021.

\bibitem{bfpcnn2}
H.~Fan, S.~Liu, Z.~Que, X.~Niu, and W.~Luk, ``High-performance acceleration of 2-d and 3-d cnns on fpgas using static block floating point,'' \emph{IEEE Transactions on Neural Networks and Learning Systems}, vol.~34, no.~8, pp. 4473--4487, 2021.

\bibitem{bfperror}
Z.~Song, Z.~Liu, and D.~Wang, ``Computation error analysis of block floating point arithmetic oriented convolution neural network accelerator design,'' in \emph{Proceedings of the AAAI Conference on Artificial Intelligence}, vol.~32, no.~1, 2018.

\bibitem{attentionisallyouneed}
A.~Vaswani, N.~Shazeer, N.~Parmar, J.~Uszkoreit, L.~Jones, A.~N. Gomez, {\L}.~Kaiser, and I.~Polosukhin, ``Attention is all you need. advances in neural information processing systems,'' \emph{Advances in neural information processing systems}, vol.~30, no. 2017, 2017.

\bibitem{stevens2021softermax}
J.~R. Stevens, R.~Venkatesan, S.~Dai, B.~Khailany, and A.~Raghunathan, ``Softermax: Hardware/software co-design of an efficient softmax for transformers,'' in \emph{2021 58th ACM/IEEE Design Automation Conference (DAC)}.\hskip 1em plus 0.5em minus 0.4em\relax IEEE, 2021, pp. 469--474.

\bibitem{zhang2022opt}
S.~Zhang, S.~Roller, N.~Goyal, M.~Artetxe, M.~Chen, S.~Chen, C.~Dewan, M.~Diab, X.~Li, X.~V. Lin \emph{et~al.}, ``Opt: Open pre-trained transformer language models,'' \emph{arXiv preprint arXiv:2205.01068}, 2022.

\bibitem{fp4}
S.-y. Liu, Z.~Liu, X.~Huang, P.~Dong, and K.-T. Cheng, ``Llm-fp4: 4-bit floating-point quantized transformers,'' \emph{arXiv preprint arXiv:2310.16836}, 2023.

\bibitem{illm}
X.~Hu, Y.~Cheng, D.~Yang, Z.~Yuan, J.~Yu, C.~Xu, and S.~Zhou, ``I-llm: Efficient integer-only inference for fully-quantized low-bit large language models,'' \emph{arXiv preprint arXiv:2405.17849}, 2024.

\bibitem{xiao2023smoothquant}
G.~Xiao, J.~Lin, M.~Seznec, H.~Wu, J.~Demouth, and S.~Han, ``Smoothquant: Accurate and efficient post-training quantization for large language models,'' in \emph{International Conference on Machine Learning}.\hskip 1em plus 0.5em minus 0.4em\relax PMLR, 2023, pp. 38\,087--38\,099.

\bibitem{frantar2022gptq}
E.~Frantar, S.~Ashkboos, T.~Hoefler, and D.~Alistarh, ``Gptq: Accurate post-training quantization for generative pre-trained transformers,'' \emph{arXiv preprint arXiv:2210.17323}, 2022.

\bibitem{liu2021post}
Z.~Liu, Y.~Wang, K.~Han, W.~Zhang, S.~Ma, and W.~Gao, ``Post-training quantization for vision transformer,'' \emph{Advances in Neural Information Processing Systems}, vol.~34, pp. 28\,092--28\,103, 2021.

\bibitem{zafrir2019q8bert}
O.~Zafrir, G.~Boudoukh, P.~Izsak, and M.~Wasserblat, ``Q8bert: Quantized 8bit bert,'' in \emph{2019 Fifth Workshop on Energy Efficient Machine Learning and Cognitive Computing-NeurIPS Edition (EMC2-NIPS)}.\hskip 1em plus 0.5em minus 0.4em\relax IEEE, 2019, pp. 36--39.

\bibitem{roundofferror}
K.~Kalliojarvi and J.~Astola, ``Roundoff errors in block-floating-point systems,'' \emph{IEEE transactions on signal processing}, vol.~44, no.~4, pp. 783--790, 1996.

\bibitem{Design1}
G.~C. Cardarilli, L.~Di~Nunzio, R.~Fazzolari, D.~Giardino, A.~Nannarelli, M.~Re, and S.~Span{\`o}, ``A pseudo-softmax function for hardware-based high speed image classification,'' \emph{Scientific reports}, vol.~11, no.~1, p. 15307, 2021.

\bibitem{design2}
Y.~Zhang, L.~Peng, L.~Quan, Y.~Zhang, S.~Zheng, and H.~Chen, ``High-precision method and architecture for base-2 softmax function in dnn training,'' \emph{IEEE Transactions on Circuits and Systems I: Regular Papers}, vol.~70, no.~8, pp. 3268--3279, 2023.

\bibitem{touvron2023llama2}
H.~Touvron, L.~Martin, K.~Stone, P.~Albert, A.~Almahairi, Y.~Babaei, N.~Bashlykov, S.~Batra, P.~Bhargava, S.~Bhosale \emph{et~al.}, ``Llama 2: Open foundation and fine-tuned chat models,'' \emph{arXiv preprint arXiv:2307.09288}, 2023.

\bibitem{meta2024llama3}
Meta, ``Meta llama 3: Advancing generative ai responsibly,'' \url{https://ai.meta.com/blog/meta-llama-3/}, 2024.

\bibitem{wiki}
S.~Merity, C.~Xiong, J.~Bradbury, and R.~Socher, ``Pointer sentinel mixture models,'' \emph{arXiv preprint arXiv:1609.07843}, 2016.

\bibitem{shao2023omniquant}
W.~Shao, M.~Chen, Z.~Zhang, P.~Xu, L.~Zhao, Z.~Li, K.~Zhang, P.~Gao, Y.~Qiao, and P.~Luo, ``Omniquant: Omnidirectionally calibrated quantization for large language models,'' \emph{arXiv preprint arXiv:2308.13137}, 2023.

\bibitem{xue2024oltron}
C.~Xue, C.~Zhang, X.~Jiang, Z.~Gao, Y.~Lin, and G.~Sun, ``Oltron: Algorithm-hardware co-design for outlier-aware quantization of llms with inter-/intra-layer adaptation,'' in \emph{Proceedings of the 61st ACM/IEEE Design Automation Conference}, 2024, pp. 1--6.

\bibitem{guo2023olive}
C.~Guo, J.~Tang, W.~Hu, J.~Leng, C.~Zhang, F.~Yang, Y.~Liu, M.~Guo, and Y.~Zhu, ``Olive: Accelerating large language models via hardware-friendly outlier-victim pair quantization,'' in \emph{Proceedings of the 50th Annual International Symposium on Computer Architecture}, 2023, pp. 1--15.

\bibitem{chisel}
J.~Bachrach, H.~Vo, B.~Richards, Y.~Lee, A.~Waterman, R.~Avi{\v{z}}ienis, J.~Wawrzynek, and K.~Asanovi{\'c}, ``Chisel: constructing hardware in a scala embedded language,'' in \emph{Proceedings of the 49th Annual Design Automation Conference}, 2012, pp. 1216--1225.

\bibitem{kurup1997logic}
P.~Kurup and T.~Abbasi, \emph{Logic synthesis using Synopsys{\textregistered}}.\hskip 1em plus 0.5em minus 0.4em\relax Springer Science \& Business Media, 1997.

\bibitem{muralimanohar2009cacti}
N.~Muralimanohar, R.~Balasubramonian, and N.~P. Jouppi, ``Cacti 6.0: A tool to model large caches,'' \emph{HP laboratories}, vol.~27, p.~28, 2009.

\bibitem{sharma2016dnnweaver}
H.~Sharma, J.~Park, E.~Amaro, B.~Thwaites, P.~Kotha, A.~Gupta, J.~K. Kim, A.~Mishra, and H.~Esmaeilzadeh, ``Dnnweaver: From high-level deep network models to fpga acceleration,'' in \emph{the Workshop on Cognitive Architectures}, 2016.

\end{thebibliography}

\end{document}